\documentclass{article}

\usepackage[preprint]{neurips_2021}

\usepackage[utf8]{inputenc} 
\usepackage[T1]{fontenc}    
\usepackage{hyperref}       
\usepackage{url}            
\usepackage{booktabs}       
\usepackage{amsfonts}       
\usepackage{nicefrac}       
\usepackage{microtype}      
\usepackage{xcolor}         

\usepackage{graphicx}
\usepackage{subfigure}

\usepackage[normalem]{ulem}
\useunder{\uline}{\ul}{}

\usepackage{cite}

\usepackage{xspace}
\usepackage{natbib}

\newcommand{\zhz}[1]{}

\makeatletter
\DeclareRobustCommand\onedot{\futurelet\@let@token\@onedot}
\def\@onedot{\ifx\@let@token.\else.\null\fi\xspace}
\def\eg{\emph{e.g}\onedot}

\def\etc{\emph{etc}\onedot}

\makeatother

\title{What does Transformer learn about source code?}

\author{%
  Kechi Zhang \\
  Peking University\thanks{Also with Key Laboratory of High Confidence Software Technologies (Peking University), Ministry of Education, China}\\
  \texttt{zhangkechi@pku.edu.cn} \\
  \And
  Ge Li  \\
  Peking University$^{\ast}$ \\
  \texttt{lige@pku.edu.cn} \\
  \And
  Zhi Jin \\
  Peking University$^{\ast}$ \\
  \texttt{zhijin@pku.edu.cn} \\
}

\begin{document}

\maketitle

\begin{abstract}
In the field of source code processing, the transformer-based representation models have shown great powerfulness and have achieved state-of-the-art (SOTA) performance in many tasks.
Although the transformer models process the sequential source code, pieces of evidence show that they may capture the structural information (\eg, in the syntax tree, data flow, control flow, \etc) as well.
We propose the aggregated attention score, a method to investigate the structural information learned by the transformer. We also put forward the aggregated attention graph, a new way to extract program graphs from the pre-trained models automatically. We measure our methods from multiple perspectives.
Furthermore, based on our empirical findings, we use the automatically extracted graphs to replace those ingenious manual designed graphs in the Variable Misuse task.
Experimental results show that the semantic graphs we extracted automatically are greatly meaningful and effective, which provide a new perspective for us to understand and use the information contained in the model.

\end{abstract}

\section{Introduction}

\zhz{Paragraph 1: How great transformer is in source code processing. Examples here.}

\zhz{Paragraph 2: Evidences of structural information learned by transformer in source code processing. Propose the problem -- Can transformer learn structural information from sequential source code?}

\zhz{Paragraph 3: Work in NLP that would supports the proposed problem. Brief introduction.}

\zhz{Paragraph 4: The proposed method. It is better to name the approach using a macro def. Describe the idea of the method.}

\zhz{Paragraph 5: The empirical results to support that transformer can capture structural information from sequence.}

\zhz{Paragraph 6: Experimental results.}

\zhz{Paragraph 7: contribution. No need to write the section structure of this paper.}

Since it was proposed, the Transformer profoundly impacts on nature language processing due to its ability of modeling long-range dependencies and its self-attention mechanism.
The last several years have witnessed that transformer-based models, such as BERT, can handle the downstream tasks with sequences. Moreover, there are also many studies on why transformer-based models can represent sequences well. 
\citep{DBLP:journals/corr/abs-1906-01698} \zhz{Do not cite like this. Use a bib file and the cite commands.} shows that BERT representations are hierarchical rather than linear. \citep{DBLP:conf/acl/WuCKL20} also shows that we can extract the global syntactic information from the transformer-based encoder. It indicates that the transformer-based models can learn about the structural information about the natural language while training on the downstream tasks.

Transformer-based models are also wildly applied in the code representation field, such as some pre-trained models, including CodeBert (\citep{DBLP:conf/emnlp/FengGTDFGS0LJZ20}) and GraphCodeBert (\citep{DBLP:journals/corr/abs-2009-08366}). Because of its complex logical structure (such as the syntax structure and control or data flow) and unique keyword distribution (including some programming idioms), the programming language is quite different from the nature language. There are more stringent rules to constrain the programming languages, while the natural language is flexible. Because of this strong structural information, in most cases, the additional structural information can effectively enhance the performance of the models in downstream code intelligence tasks.
It raises an emerging question for us that what Transformer learns about source code. However, most previous model explanation methods in natural language fail on revealing the special structural information of the source code learned by Transformer. It is important to propose an interpretation method suitable for program language. Solving this question can guide us to use the transformer-based models to represent the code effectively and help us explore the ability of models to learn structural information.

Our experiments are based on these thinkings. We propose the aggregated attention score, a new method to capture the structural information learned by the transformer-based model. We analyze the aggregated attention score of each layer and summarize an effective way to extract structural information.
Considering that the process of the existing manual construction of program representation graph is very complex, sometimes it needs to define many rules and use specific tools, we also propose the aggregated attention graph, a method to automatically extract the program representation graph from pre-trained transformer-based models like CodeBert and GraphCodeBert with the Chu-Liu/Edmonds (CLE) algorithm (\citep{1965On}) and attention head information. We evaluate the obtained graph quantitatively and apply it to a specific downstream task. The results not only show that automatically extracted graphs can maintain a comparable performance compared with the ingenious manual designed graph, but also proves the effectiveness of our analysis method. Through these experiments, We provide a new perspective to understand and utilize the structural information Transformer learns about source code.

The contributions of our work are as follows:
\begin{itemize}
\item We introduce a new method, the aggregated attention score, to capture the structural information learned by the transformer-based model. We take CodeBERT and GraphCodeBert as examples. 
\item We also propose the aggregated attention graph, an effective way to automatically extract the program representation graph from pre-trained transformer-based models. We put forward some methods to measure the obtained graphs in terms of quantity and show the rich semantic information in the graphs.
\item We apply the aggregated attention graph to the Variable Misuse task with the state-of-the-art model, Graph Relational Embedding Attention Transformers (GREAT for short) (\citep{DBLP:conf/iclr/HellendoornSSMB20}). The result shows that our work can maintain a comparable performance with smaller graphs compared with the manual designed graphs, proving the correctness of our analysis method. It offers an insight to understand and utilize the structual information learned by Transformer about source code.
\end{itemize}


\zhz{Related Work here. This paper needs many more citations. I list some possible related work here.}

\zhz{1. Source code processing. 2. Model explanation.}

\section{Background}

\subsection{Transformer}

\zhz{You need give the readers an overview of the transformer model. Like... what is it? or Why you have to introduce it here. or Why do you put some equations here (maybe a figure is more easy to understand?). You must lure the readers. The current version is to force them to catch up with you..}


Transformer (\citep{DBLP:conf/nips/VaswaniSPUJGKP17}) is a model architecture relying on the attention mechanism. Given input tokens $\{x_i\}_{i=1}^{\left | x \right | }$, we pack their word embeddings to a matrix $X^0 = \left [ x_1,\dots ,x_{\left | x \right | } \right ] $. The stacked $L$-layer Transformer computes the final output via $X^{l} = Transformer_l(X^{l-1}), l \in \left [1,L \right ]$.
The core component of a Transformer block is multi-head self-attention. The $h$-th self-attention head is described as:
\begin{equation}
Q_{h}=X W_{h}^{Q}, K=X W_{h}^{K}, V=X W_{h}^{V}
\end{equation}
\begin{equation}
A_{h}=softmax\left(\frac{Q_{h} K_{h}^{\top}}{\sqrt{d_{k}}}\right)
\end{equation}
\begin{equation}
H_{h}=AttentionHead \left(X\right) = A_{h}V_{h}
\end{equation}

where $Q,K \in {\mathbb{R}}^{n \times d_k}, V \in {\mathbb{R}}^{n \times d_v} $, and the score $A_{i,j}$ indicates how much attention token $x_i$ puts on $x_j$. There are usually multiple attention heads in a Transformer block. The attention heads follow the same computation despite using different parameters. Let $\left | h \right |$ denote the number of attention heads in each layer, the output of multi-head attention is given by $MultiHead(X) = \left [ H_1,\dots, H_{\left | h \right |} \right ]W^{o}$, where $W^{o} \in {\mathbb{R}}^{\left | h \right |d_v \times d_x}$, $\left [ \dots \right ]$ means concatenation, and $H_i$ is computed as in Equation (3).

\subsection{CodeBERT}

\zhz{Also, CodeBERT here has a similar problem. Why do you introduce CodeBERT here? What is good about it? Any figures for the readers to understand?}

In our experiments, we employ the CodeBERT (\citep{DBLP:conf/emnlp/FengGTDFGS0LJZ20}).
CodeBERT is the first large NL-PL pre-trained model. It is a bimodal pre-trained model based on Transformer with 12 layers, 768 dimensional hidden states, and 12 attention heads for programming language (PL) and natural language (NL) by masked language modeling and replaced token detection to support text-code tasks such as code search. 

\subsection{GraphCodeBERT}
In our experiments, we also employ the GraphCodeBert model (\citep{DBLP:journals/corr/abs-2009-08366}).
GraphCodeBERT is a Transformer-based pre-trained model for programming language that incorporates data flow information in the graph representation of variables in the code. The data flow graph encodes the structure of variables based on “where-the-value-comes-from” from the AST parse. The pre-trained model is jointly trained on the code, the natural language comment of the code, and the data flow graph of the code. GraphCodeBERT includes 12 layers of Transformer with 768-dimensional hidden states and 12 attention heads. With additional structure information added to the attention score, GraphCodeBert can deliver improvements on Natural Language Code Search, Code Clone Detection, Code Translation, and Code Refinement.

Both the CodeBert model and the GraphCodeBert we use are from the official repository\footnote{\url{https://github.com/microsoft/codebert}}, which use the huggingface/transformers framework.

\section{Methods: Aggregated Attention Score and Graph}

\zhz{Do not name a section with ``methods''. Use the name of the proposed method}
\paragraph{Aggregated Attention Score}
\citep{DBLP:journals/corr/abs-1906-01698} shows that BERT representations are hierarchical rather than linear. There is something akin to the syntactic tree structure in addition to the word order information. It suggests that we can extract syntactic information of the input sequence from the transformer-based model. Due to the powerful multi-head self-attention mechanism, we have designed a method to extract information from the attention weights generated by Transformer. CodeBERT consist of 12 transformer layers, and the number of attention heads in each layer $\left | H \right | = 12$. In order to aggregate attention scores among all attention heads, we define \textbf{the aggregated attention score}, which takes the maximum over different heads for each layer:
\begin{equation}
\label{AggregatedAttn_eq}
AggregatedAttn_{l}[i,j] = \max_{h=1}^{H}h_{l}^{h}[i,j] 
\end{equation}

where the score $h_{l}^{h}[i,j]$ represents how much attention token $x_i$ puts on $x_j$ in the $h$-th self-attention head in the $l-th$ layer. 

\begin{figure}[htb]

  \centering
    \includegraphics[scale=0.5]{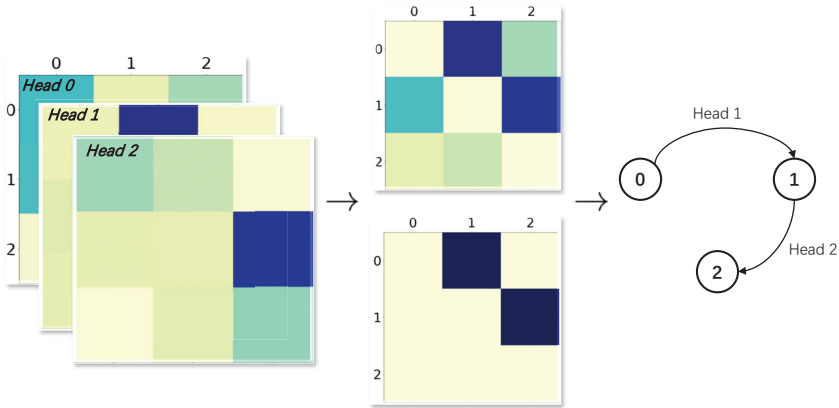}
  \caption{Illustration of the Aggregated Attention Graph. The maximum spanning graph is extracted from the aggregated attention score, with edge types from attention head information. All reflexive edges are removed. }
  \label{extraction_architecture}
\end{figure}

\paragraph{Aggregated Attention Graph}
In order to make the matrix sparse and extract hierarchical structure information, we regard the aggregated attention score as the edge weight between tokens and apply the Chu-Liu-Edmonds algorithm to find the maximum spanning tree. Then we can get \textbf{the aggregated attention graph}, a syntactic graph learned by each layer in the transformer-based model.

Furthermore, through our analytical experiments in section \ref{head_specific_relation_section}, we find that different attention heads have a different semantic preference when extracting information. It provides natural edge type information for us while constructing semantic graphs automatically without any manual definition rules. Finally, we can extract a programming representation graph with abundant edge types which contain rich semantic information. The whole process of the automatic program graph extraction is shown in Figure \ref{extraction_architecture}.

\section{Analytical Experiments}
\label{analytical_experiments}

We employ the CodeBERT model and GraphCodeBERT model and conduct several analytical experiments. 
In order to find the most suitable model for program graph extraction, we also consider the effect of fine-tuning on the pre-trained transformer-based models. We fine-tune models in Code Summarization task (\citep{DBLP:conf/acl/IyerKCZ16}) using the Python dataset from CodeSearchNet dataset (\citep{husain2019codesearchnet,DBLP:journals/corr/abs-2102-04664}) guided by the official repository.\footnote{\url{https://github.com/microsoft/CodeXGLUE}}
The objective of the Code Summarization task is to generate the natural language comment for a code, which requires reasoning on the semantics of complete code snippets. 

Based on these original or fine-tuned pre-trained models,
we analyze the aggregated attention score from different layers and find that the structural information can be extracted from the deepest layer. We use the proposed graph extraction method to obtain the representation graph from attention heads and try to evaluate the graph quantitatively. We also find that different attention heads have different semantic preferences, which provides nature edge types for our proposed extraction method. 

All the input sample code comes from the CodeSearchNet dataset. For the convenience of experiments, we choose the part of Python because it accounts for the largest proportion of the six languages, and there are many parsing utilities for Python source codes.

\subsection{Attention Patterns Analysis between layers}
\label{attention_pattern_section}
\begin{figure}[htb]

  \centering
    \includegraphics[scale=0.25]{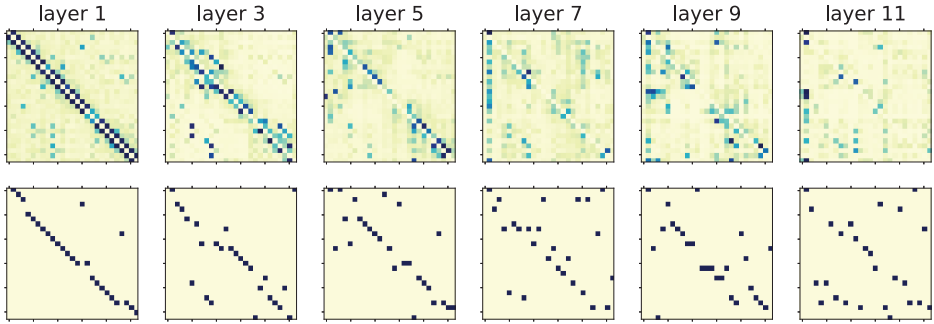}
  \caption{Illustration of the aggregated attention scores and the maximum spanning tree from different layers in CodeBERT. As the layers deepen, the aggregated attention scores change from diagonal pattern to heterogeneous pattern.}
  \label{attns_layer}
\end{figure}

Using the Equation(\ref{AggregatedAttn_eq}), we calculate the aggregated attention score for each layer given an input sequence. In figure \ref{attns_layer} we show the aggregated attention scores and their corresponding maximum spanning tree between different layers given an example input in CodeBERT. We notice that similar findings can be observed in GraphCodeBERT. As the layers deepen, the aggregated attention scores change from diagonal patterns to heterogeneous patterns. The diagonal pattern means the attention weights focus on neighbor words. In contrast, the heterogeneous pattern means the attention weights become more focused on the semantically related words but far apart. It stands to reason that the lower layers have the most information about linear word order, and the higher layers may have more information about semantic knowledge and task-specific knowledge. 

It hints that we can extract syntactic information from the attention heads in deep layers. Figure \ref{attns_layer} also shows that the attention score matrix is too dense to analyze, especially in deep layers. Using the maximum spanning tree algorithm is an effective method for extracting features from aggregated attention scores more clearly.


\subsection{Effectiveness Analysis of Graphs}

\begin{figure}[htb]
  \centering

\subfigure[An example of the aggregated attention graph] {
 \label{codebert_tree}     
\includegraphics[scale=0.4]{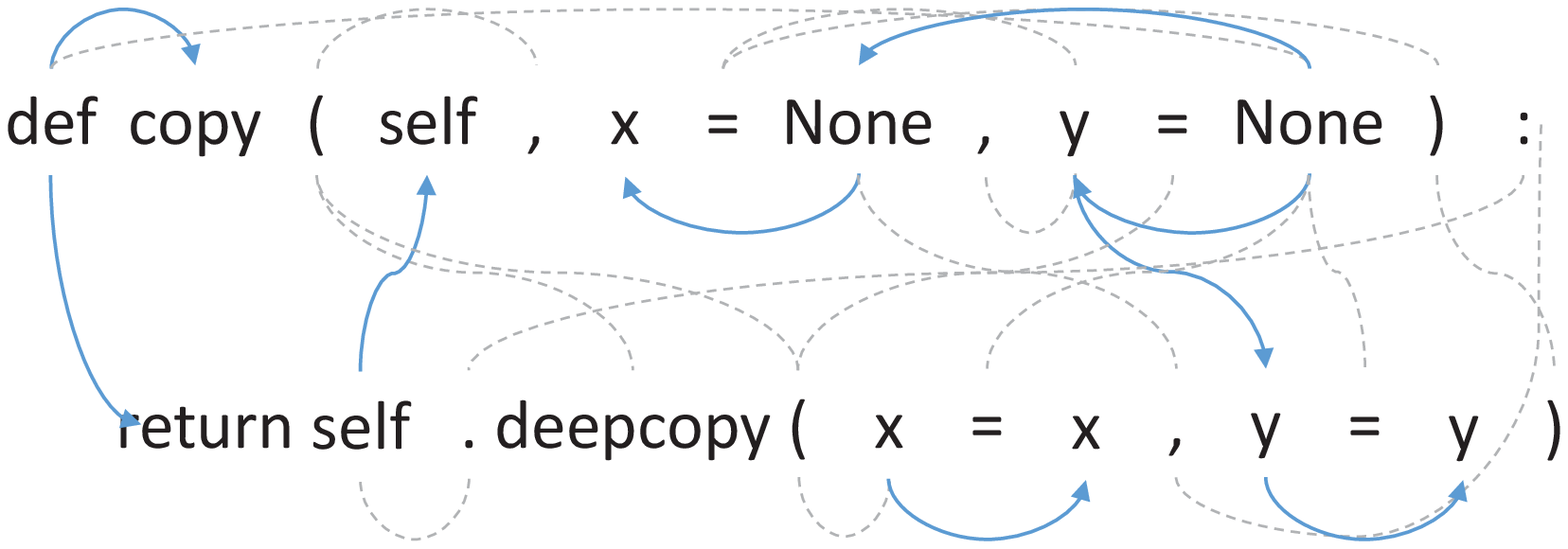}
}
\subfigure[Corresponding concrete syntax tree(CST)] {
 \label{tree_sitter}     
\includegraphics[scale=0.8]{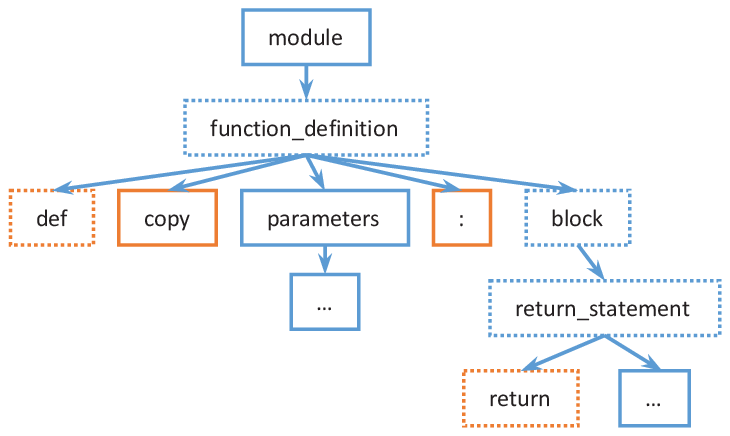}
}
    \caption{Illustration of (a): an example of the aggregated attention graph. Each line represents an edge between two tokens. We filter out some edges with obvious semantic features and mark them with blue lines. (b):Corresponding concrete syntax tree(CST). Some nodes are simplified. The intermediate nodes between \textit{def} and \textit{return} are displayed in blue dotted borders.}
\end{figure}

Based on our finding in section \ref{attention_pattern_section}, we use the deepest aggregated attention scores to extract semantic knowledge. Figure \ref{codebert_tree} displays a aggregated attention graph induced from the aggregated attention score in the last layer in CodeBERT with the maximum spanning tree algorithm. We can find that a lot of important structural relations, including control flows and data flows, can be extracted from the attention scores, including the relation between keywords \textit{def} and \textit{return}, the relation between the two sides of an assignment statement, the relation between the definition of a variable and its usage, \etc. It suggests that transformer-based models can learn complex structural information from the source code through the self-attention mechanism.

\begin{figure}[htb]
  \centering

\subfigure[Distribution of the tree distance] {
\includegraphics[scale=0.4]{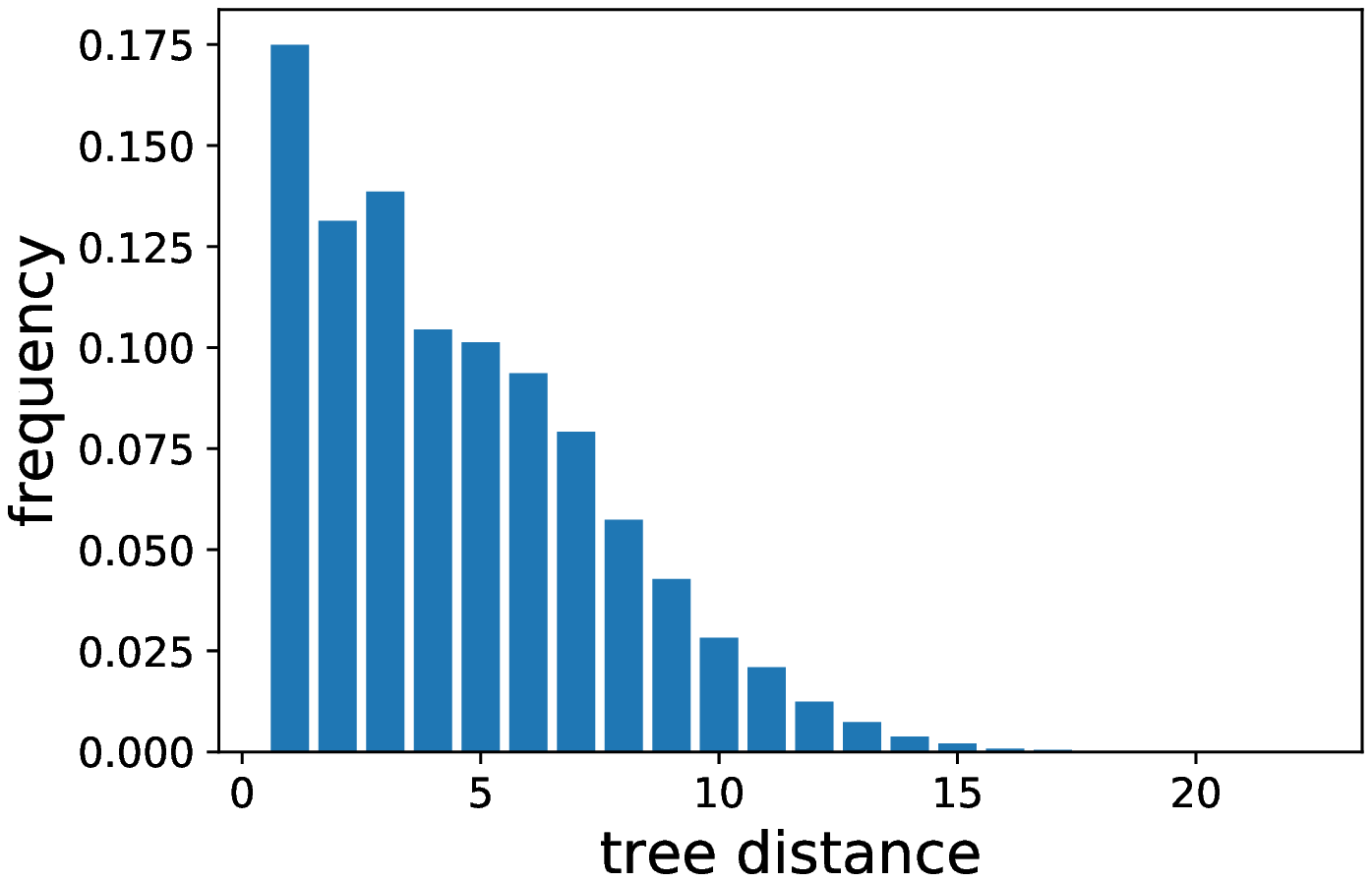}
}
\subfigure[Relationship between tree distance and sequence distance] {
 \label{count_tree_sitter_tree_seq}     
\includegraphics[scale=0.4]{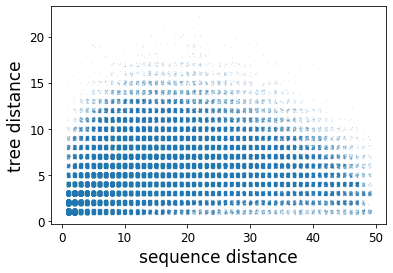}
}
    \caption{Illustration of (a): the distribution of the distance in the concrete syntax tree for our extracted tree in fine-tuned CodeBERT, and (b): the relationship between the tree distance and the sequence distance in fine-tuned CodeBERT. The deep color indicates a high frequency corresponding to the point.}
  \label{count_tree_sitter}
\end{figure}

In order to provide a shred of subjective evidence for quantitative evaluation, we compare the aggregated attention graphs with the concrete syntax tree. We define the distance between tokens in the concrete syntax tree as follows: given two tokens, \textbf{the tree distance} is the number of intermediate nodes on the shortest path between them.
For example, the tree distance between \textit{def} and \textit{return} in Figure \ref{tree_sitter} is 3 where all the intermediate nodes are displayed in blue dotted borders. A short tree distance means that tokens are semantically closely related.
Take the fine-tuned CodeBERT as an example; we show our statistical results of the tree distance in Figure \ref{count_tree_sitter}.
We conduct the same statistical experiments on original CodeBERT and GraphCodeBERT models and have observed similar findings.

Considering that neighbor words also lead to short tree distances, which are less meaningful than those that are far apart but closely related semantically, we also count the distance on the sequence. Figure \ref{count_tree_sitter_tree_seq} shows that no matter how far two tokens are in the sequence, we can still extract a relation from the attention scores if they are very close in the concrete syntax tree. This situation often happens in programming languages. For example, given a simple source code with keywords \textit{if} and \textit{else}, these two keywords may be far apart in sequence but close in the concrete syntax tree. By the way, this kind of situation appears less in natural language, especially in a simple sentence.

\begin{figure}[htb]
  \centering

\subfigure[Distribution in original CodeBERT] {
 \label{count_parent_original_codebert}     
\includegraphics[scale=0.38]{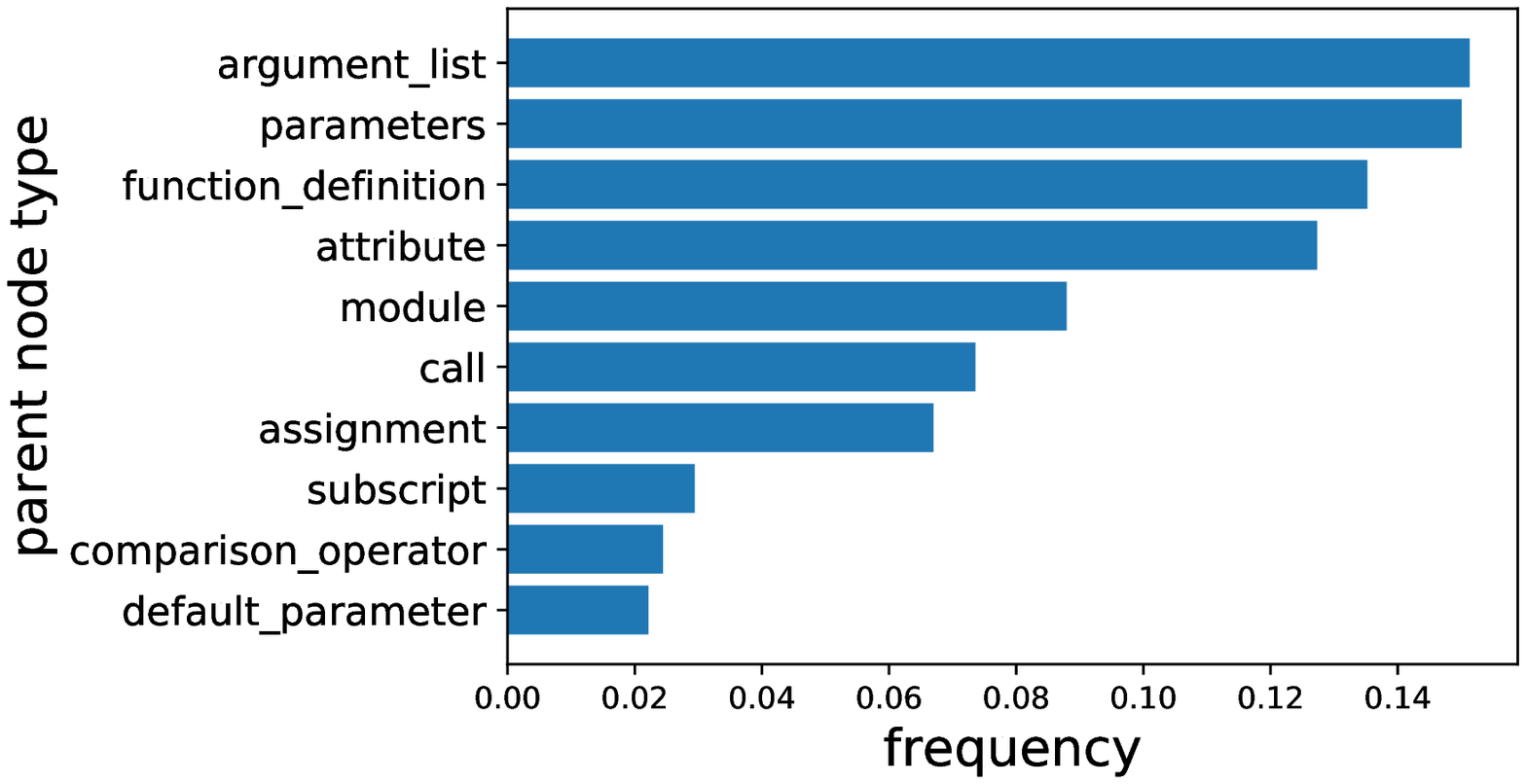}
}
\subfigure[Distribution in fine-tuned CodeBERT] {
 \label{count_parent_finetune_codebert}     
\includegraphics[scale=0.38]{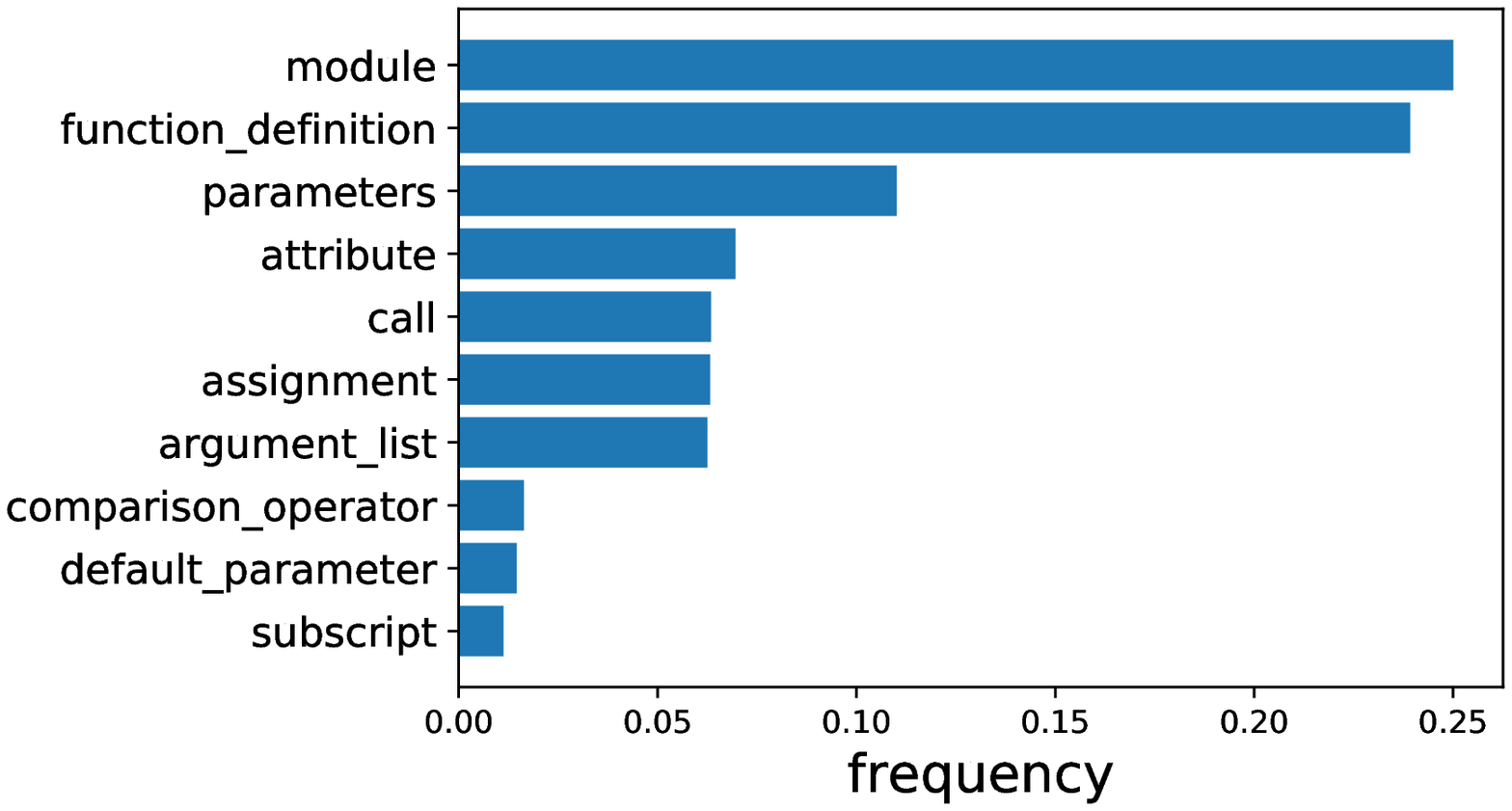}
}
    \caption{Illustration of (a)(b): last common parent node type distribution in original CodeBERT and fine-tuned CodeBERT. We show the 10 most frequent parent node types in the figure.}
  \label{count_parent_codebert}
\end{figure}

We also find the differences between original models and fine-tuned models. By getting the last common parent node in concrete syntax trees, We explore the factual semantic information in the edge of the aggregated attention graph. Statistical results are shown in Figure \ref{count_parent_codebert}. We find that the original models prefer to capture common semantic relations across different programming languages, including the relations between adjacent parameters or arguments. The fine-tuned models prefer to capture some language-specific structural information. For example, models fine-tuned in Python dataset are sensitive to the relation about the language-specific function definition forms, such as \textit{def} and \textit{return}. We sample 1000 source code examples and find 122 \textit{def-return} edges in graphs extracted from fine-tuned CodeBERT. But for the original CodeBERT we only find 9 \textit{def-return} edges. This finding 
proves that fine-tuning on specific programming language can help pre-trained models learn the language-specific structural information, which is instructive for extracting graphs automatically from transformer-based models.



\subsection{Head-specific Relation Analysis}
\label{head_specific_relation_section}
\begin{figure}[htb]
  \centering

\subfigure[\textit{self-self} edge] {
 \label{self_self_edge}     
\includegraphics[scale=0.4]{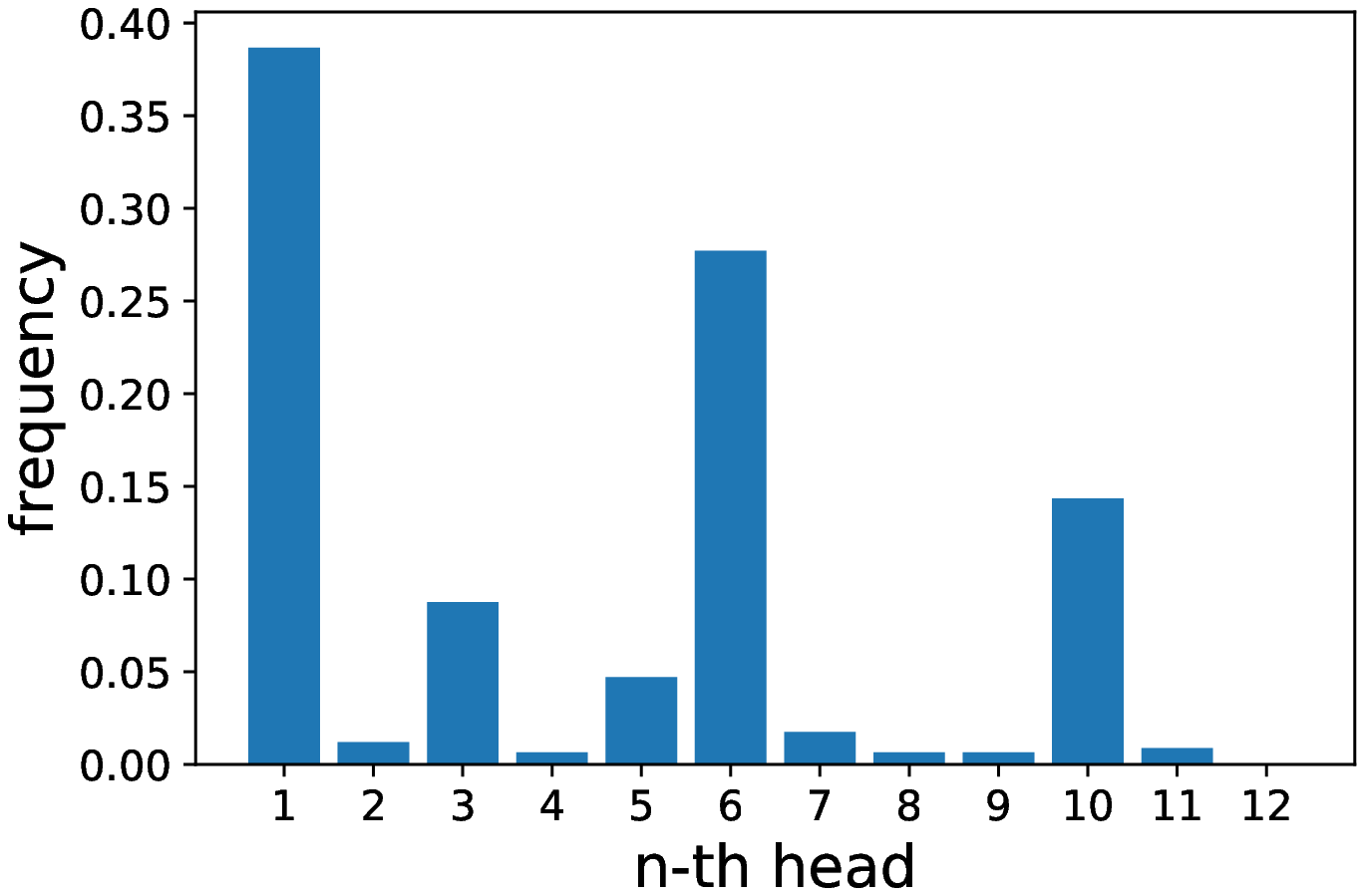}
}
\subfigure[\textit{def-return} edge] {
 \label{def_return_edge}     
\includegraphics[scale=0.4]{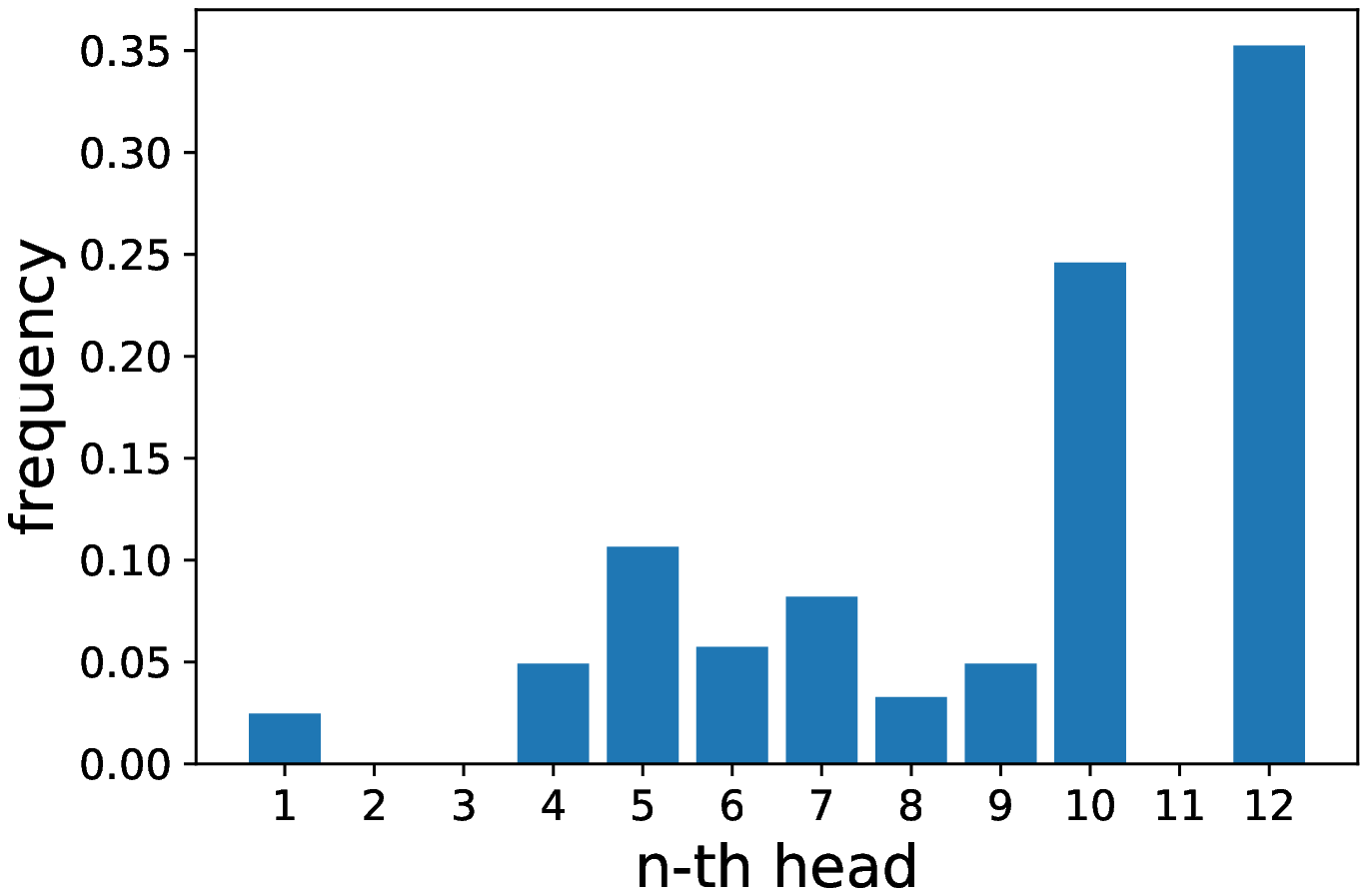}
}
    \caption{The distribution of two different edges on attention heads in fine-tuned CodeBERT}
  \label{count_edge_head}
\end{figure}

Besides, we analyze the relationship extracted from different attention heads. We sample 1000 source code examples and extract their syntactic graph from the aggregated attention scores in the last layer. Specially we collect 1826 edges connecting word \textit{self} in different positions and 122 edges connecting word \textit{def} and word \textit{return}. We count the attention head to which each edge belongs, and statistical results are shown in Figure \ref{count_edge_head}. 

The \textit{self-self} edge is closely related to data flow and appears more in the $1st$ head, the $6th$ head, and $10th$ head. Correspondingly, the \textit{def-return} edge is closely related to the code structure and appears more in the $10th$ head and the $12th$ head. Similar findings have been found in other models, confirming that different attention heads have a different semantic preference when extracting information. It also proves the effectiveness of the multi-head attention mechanism when extracting complex structural information.

\section{Applications: VarMisuse}
\label{application}

Our analytical experiments find that it is possible to automatically extract representation graphs from pre-trained transformer-based models. We also find that fine-tuning can help the model learn language-specific structural information better. 

In order to further prove the effectiveness and correctness of our methods, we choose the Variable Misuse task (\citep{DBLP:conf/iclr/AllamanisBK18,DBLP:conf/iclr/VasicKMBS19}) as the application experiment of our proposed method. The state-of-the-art model is the Graph Relational Embedding Attention Transformer (GREAT for short) (\citep{DBLP:conf/iclr/HellendoornSSMB20}), which uses the program representation graph as inputs. The dataset for the Variable Misuse task used by the GREAT model does not release the construction tools, making it hard to use on other datasets. In our application experiments, we will replace the original input graphs with our aggregated attention graphs.

\subsection{Experimental Setup}
\label{section_experiment_setup}


\paragraph{The VarMisuse Task}
Detecting variable misuses in code is a task that requires understanding and reasoning about program semantics. We follow the previous work (\citep{DBLP:conf/iclr/AllamanisBK18,DBLP:conf/iclr/VasicKMBS19})
and focus on the localization-and-repair task.
The task is defined as follows: Given a function, predict two pointers into the function’s tokens, one pointer for the location of a variable use containing the wrong variable (or a special no-bug location), and one pointer for any occurrence of the correct variable that should be used at the wrong location instead.

\paragraph{Model}
\citep{DBLP:conf/iclr/HellendoornSSMB20} has proposed the Graph Relational Embedding Attention Transformers(GREAT for short), which bias traditional Transformers with relational information from graph edge types, and have achieved good performance with their ingenious design of graph representation of code. The graph includes control flow edges, data flow edges, semantic edges, syntactic edges, which takes many human resources to define and implement these rules.

In the experiments, the GREAT model we use is from the official repository.\footnote{\url{https://github.com/VHellendoorn/ICLR20-Great}} 
The number of layers is 6, and the attention dimension is set to 256. 
All the experiments are conducted on two Tesla V100 GPUs. 

\paragraph{Metrics}
We measure four accuracy metrics in the experiments to fully analyze the performance of different input graphs trained with the GREAT model with the same parameter settings.
All accuracies are reported as: joint localization \& repair accuracy (the key indicator, for buggy samples, how often the model correctly localizes and repairs the bug), bug-free classification accuracy (whether the model correctly identifies the method as (non-)buggy), bug-localization accuracy (whether the model correctly identifies the bug’s location for buggy samples) and repair accuracy (whether the model points to the correct variable to repair the bug).

\paragraph{Dataset}

This dataset provided by the official repository \footnote{\url{https://github.com/google-research-datasets/great}} is generated synthetically from the corpus of Python code in the ETH Py150 Open dataset (\citep{DBLP:conf/oopsla/RaychevBV16}). While the construction tools are not provided, the dataset provides the source tokens, which can be utilized by our automatic program graph extraction method.


\subsection{Input Graph Analysis}

Based on our finding that fine-tuning on specific programming language can help pre-trained models learn the language-specific structural information, we use the fine-tuned CodeBERT and GraphCodeBERT 
in Code Summarization task with Python dataset in our experiments.
We use the aggregated attention graphs extracted from these models to replace the original rule-based graphs, respectively.
We also notice that there are some special symbols for representing formats such as \textit{\#NEWLINE\#} that are not in the vocabulary of CodeBERT. We replace them with \textit{<mask>} while extracting semantic graphs and do not record the edges on them, which makes our graph smaller and more reasonable. Because the graph is too small and the aggregated attention scores in the deep layer may lose information about linear word order as described in section \ref{attention_pattern_section}, we add sequence edges as a new edge type in our graph.

In addition, we also use a sequence structure graph, which only connects adjacent words, as our baseline. Statistics of these graphs are listed in table \ref{varmisuse_graph}, and the statistics of edge types include the reverse edges.

 \makeatletter\def\@captype{table}\makeatother
\begin{minipage}[t]{0.4\textwidth}

  \caption{Statistics of our experiment graphs}\label{varmisuse_graph}
  \centering
  \begin{tabular}{lll}
    \toprule
             & \begin{tabular}[c]{@{}l@{}}Edge\\ types\end{tabular}      & \begin{tabular}[c]{@{}l@{}}Avg.\\ edges\end{tabular} \\
    \midrule
    Origin & 22  & 193.3     \\
    Sequence     & 2 & 114.9   \\
    \textit{Ours}       \\
    CodeBERT     & 26       & 185.4  \\
    GraphCodeBERT     & 26       & 183.2  \\
    \bottomrule
  \end{tabular}

\end{minipage}
 \makeatletter\def\@captype{figure}\makeatother
 \hfill
\begin{minipage}[t]{0.5\textwidth}
  \caption{The proportion of edge types in original graphs extracted from CodeBERT}
     \label{great_edge_type}
    \includegraphics[width=\textwidth]{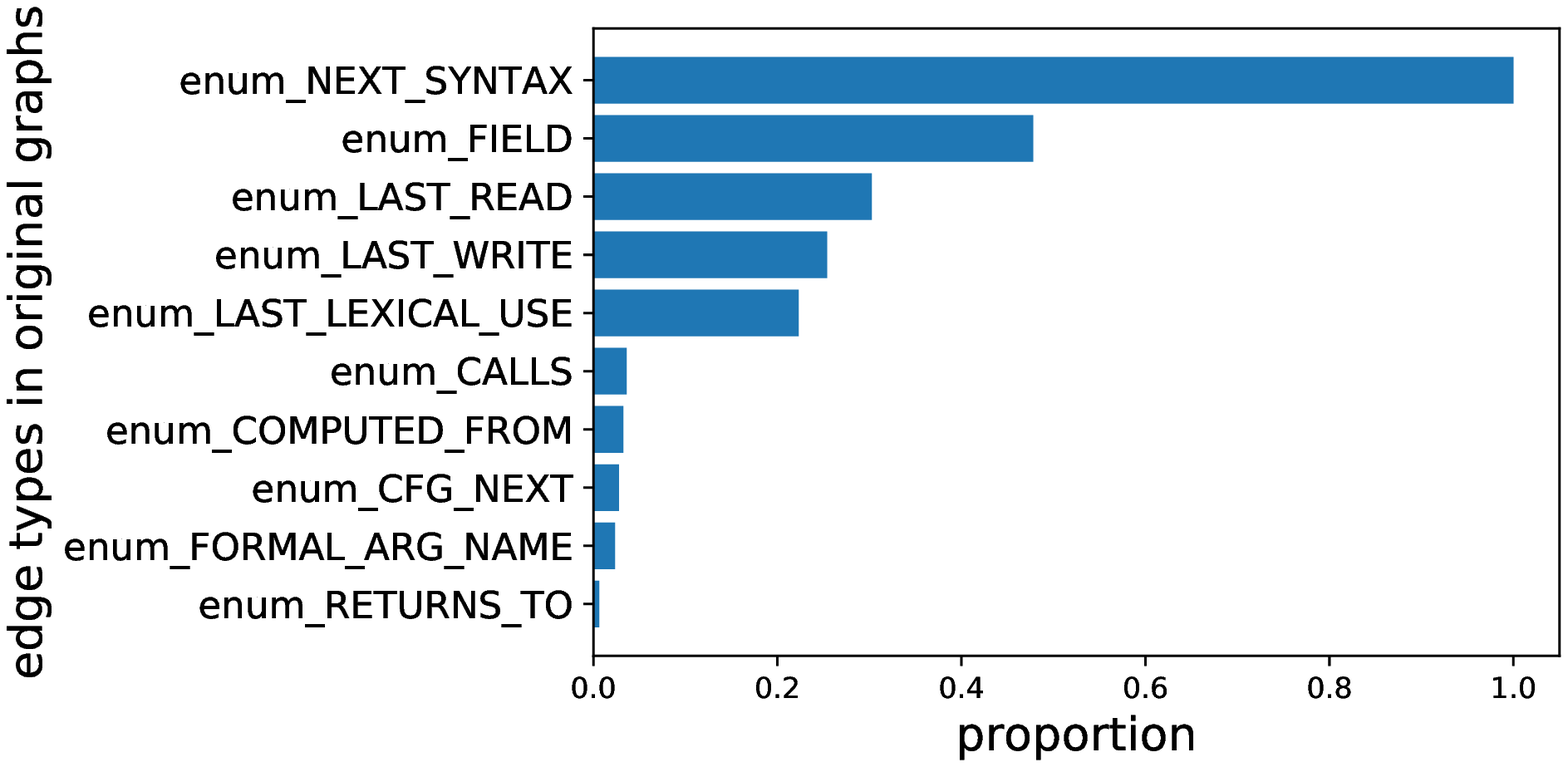}

\end{minipage}

We also measure the degree of coincidence between our constructed graphs and the original graphs. There are 22 edge types in original graphs, including semantic edges, syntactic edges, control-flow edges, and data flow edges. All these edges are generated by ingenious manual design. We analyze the automatically extracted graph and explore the proportion of each edge type in the original graph in Figure \ref{great_edge_type}. We find that except for \textit{NEXT\_SYNTAX} edge, there are many syntactic edges (\textit{FIELD}, \textit{LAST\_LEXICAL\_USE}) and data flow edges (\textit{LAST\_READ}, \textit{LAST\_WRITE}) that can be extracted from the pre-trained models automatically.

The statistical results mean that without any artificial parsing rules when constructing graphs, our graphs are still quite similar to the ingenious manual design. Moreover, we further explore the unique edges in our graphs and find some edges connected to brackets and other symbols. These edges can not be well explained in current artificial rules, but their effectiveness is still worth exploring in future work.

It also hints that the self-attention mechanism can well handle some implicit syntactic and data flow relationships. However, it is still limited to complex relationships such as control flows.
In the Variable Misuse task, it is obvious that the data flow relations play an important role in the model's performance, which is consistent with the feature of the graph we constructed.
\subsection{Results}

The experiment results are shown in table \ref{varmisuse_rst}.
We are surprised to find that even though the CodeBERT model and the GraphCodeBERT model we used to extract relations are only fine-tuned on the Code Summarization task, the graphs we extracted are still very effective and meaningful. Although the final result drops slightly compared with the original graphs, we can still conclude that the aggregated attention graph we extracted contains a lot of real syntax information and can be utilized in downstream tasks. The advantage of our graph is that it is smaller and needs no manual design.

We also notice that with the same training steps, the graphs from GraphCodeBERT slightly outperform those from CodeBERT. We think that using the code structure information during pre-training is essential, but the fine-tuning process may weaken this advantage. However, unfortunately, because the original pre-trained model is relatively insensitive to the language-specific structural information, it is hard for us to explore it further.
\begin{table}[htb]
\caption{Test-set results of our experiment graphs with GREAT model on VarMisuse task}\label{varmisuse_rst}
\centering
\begin{tabular}{lllll}
\hline
                        & Loc \& Rep Acc & Class. Acc & Loc Acc & Rep Acc \\ \hline
origin                       & 78.18\%                           & 90.36\%             & 86.16\% & 85.62\% \\
sequence                     & 72.12\%                           & 91.99\%             & 79.09\% & 81.77\% \\
\multicolumn{5}{l}{{\textit{ours}}}  \\
from CodeBERT                      & 77.33\%                           & 88.98\%             & 85.77\% & 84.38\% \\ 
from GraphCodeBERT                        & 77.81\%                           & 90.38\%             & 85.85\% & 85.37\% \\ \hline
\end{tabular}
\end{table}

The positive results on the VarMisuse task prove our point that transformer-based models can extract rich syntactic information with the self-attention mechanism, and this syntactic information is available in downstream tasks. Our method of automatically extracting program graphs from the aggregated attention score is effective. It provides a new perspective for us to understand and use the information contained in the model. Furthermore, we can continue to mine the useful information in attention scores, especially those relationships that cannot be well explained in current artificial rules.

\section{Related Work}



There has been substantial research investigating what pre-trained transformer models have learned about nature languages’ structures.

\citep{DBLP:journals/corr/abs-1906-01698} shows that BERT representations are hierarchical rather than linear.
\citep{DBLP:conf/emnlp/KovalevaRRR19} proposes five attention patterns and finds the "heterogeneous" attention pattern could potentially be linguistically interpretable.
\citep{DBLP:journals/corr/abs-1906-04341} lists many examples to explore attention scores in BERT (\citep{DBLP:conf/naacl/DevlinCLT19}) and thinks that it is possible to extract parse trees from BERT heads, and several studies focused on it.
\citep{DBLP:conf/acl/JawaharSS19} includes a brief illustration of a dependency tree extracted directly from self-attention weights. 
\citep{DBLP:journals/corr/abs-1911-12246} uses two simple dependency relation extraction methods and finds neither of their analysis methods supports the
existence of generalist heads that can perform holistic parsing.
\citep{DBLP:journals/corr/abs-1906-01958} proposes a transparent deterministic method of
quantifying the amount of syntactic information present in the self-attentions.
\citep{DBLP:journals/corr/abs-2004-11207} proposes a self-attention attribution method (ATTATTR) based on integrated gradient to interpret the information interactions inside Transformer.

Recently there are some work argues that self-attention distributions are not directly interpretable in nature language (\citep{DBLP:conf/acl/SerranoS19,DBLP:conf/naacl/JainW19,DBLP:conf/iclr/BrunnerLPRCW20}). \citep{DBLP:journals/tacl/RogersKR20} reviews the current state of knowledge about how BERT works and hold the opinion that the syntactic structure of nature language is not directly encoded in self-attention weights. Many recent approaches use the feature representations (\citep{DBLP:conf/naacl/HewittM19,DBLP:conf/nips/ReifYWVCPK19,DBLP:journals/corr/abs-1906-11511}) generated by a pre-trained model and train a small model to perform a supervised task (e.g., dependency labeling). 
Most recently, \citep{DBLP:conf/acl/WuCKL20}
shows that we could use encoder representations instead of attention scores to recover syntactic trees of the nature language from BERT. We have tried this method and found that it does not improve extracting deep semantic information of source code.

We hypothesize that there are more complex logical structures in the source code. For example, in source code, there may be two words that are far apart but closely related semantically, such as \emph{if} and \emph{else}. However, it is less likely to happen in natural language, especially in a simple sentence. Because of these structural differences, many methods of interpreting natural language models are not ideally suited to programming languages. The interaction between words may be more effective for information extraction in source code, just like the key-value attention mechanism.

\section{Conclusion}

In this paper, we try to figure out what transformer-based models learn about source code by proposing a new method, the aggregated attention score. Based on our method, we put forward the aggregated attention graph, a new way to automatically extract representation graphs from pre-trained models. We measure our methods from multiple perspectives and apply our methods on VarMisuse task. The results also prove the correctness of our methods and analytical experiments.

In the future, we plan to further explore other semantic information extracted from transformer-based models. We would also rethink the implicit information in pre-trained transformer models and propose new models to fully utilize these task-specific features in different code intelligence tasks.






\bibliographystyle{unsrtnat}
\bibliography{reference}

\section*{Checklist}


\begin{enumerate}

\item For all authors...
\begin{enumerate}
  \item Do the main claims made in the abstract and introduction accurately reflect the paper's contributions and scope?
    \answerYes{}
  \item Did you describe the limitations of your work?
    \answerYes{}
  \item Did you discuss any potential negative societal impacts of your work?
    \answerNA{}
  \item Have you read the ethics review guidelines and ensured that your paper conforms to them?
    \answerYes{}
\end{enumerate}

\item If you are including theoretical results...
\begin{enumerate}
  \item Did you state the full set of assumptions of all theoretical results?
    \answerNA{}
	\item Did you include complete proofs of all theoretical results?
    \answerNA{}
\end{enumerate}

\item If you ran experiments...
\begin{enumerate}
  \item Did you include the code, data, and instructions needed to reproduce the main experimental results (either in the supplemental material or as a URL)?
    \answerYes{}
  \item Did you specify all the training details (e.g., data splits, hyperparameters, how they were chosen)?
    \answerYes{See Section~\ref{section_experiment_setup}.}
	\item Did you report error bars (e.g., with respect to the random seed after running experiments multiple times)?
    \answerNA{}
	\item Did you include the total amount of compute and the type of resources used (e.g., type of GPUs, internal cluster, or cloud provider)?
    \answerYes{See Section~\ref{section_experiment_setup}.}
\end{enumerate}

\item If you are using existing assets (e.g., code, data, models) or curating/releasing new assets...
\begin{enumerate}
  \item If your work uses existing assets, did you cite the creators?
    \answerYes{}
  \item Did you mention the license of the assets?
    \answerNA{}
  \item Did you include any new assets either in the supplemental material or as a URL?
    \answerNA{}
  \item Did you discuss whether and how consent was obtained from people whose data you're using/curating?
  \answerNA{}
  \item Did you discuss whether the data you are using/curating contains personally identifiable information or offensive content?
    \answerNA{}
\end{enumerate}

\item If you used crowdsourcing or conducted research with human subjects...
\begin{enumerate}
  \item Did you include the full text of instructions given to participants and screenshots, if applicable?
    \answerNA{}
  \item Did you describe any potential participant risks, with links to Institutional Review Board (IRB) approvals, if applicable?
    \answerNA{}
  \item Did you include the estimated hourly wage paid to participants and the total amount spent on participant compensation?
    \answerNA{}
\end{enumerate}

\end{enumerate}





\end{document}